# TOMOGRAPHY OF COLLISIONLESS STELLAR SYSTEMS

L. CIOTTI
*Osservatorio Astronomico di Bologna,*
*via Zamboni 33, CP596, I-40100 Bologna, Italy*

**Abstract.**
 In this paper the concept of tomography of a collisionless stellar system of general shape is introduced, and a generalization of the Projected Virial Theorem is obtained. Applying the tomographic procedure we then derive a new family of virial equations which coincides with the already known ones for spherically symmetric systems. This result is obtained without any use of explicit expressions for the line-of-sight velocity dispersion, or spherical coordinate system.

## 1 Introduction

A very important problem of stellar dynamics is the mass determination of stellar systems using their line-of-sight (l.o.s.) velocity dispersion $\sigma_{los}^2$ and surface brightness profile $I$. A useful mass estimator, based on the second order Projected Virial Theorem (PVT) for spherical collisionless stellar systems, is obtained integrating over the whole projection plane $\sigma_{los}^2$ using $I$ as weight function, (cf. e.g., Ciotti and Pellegrini, 1992). Indeed the classical PVT (see, e.g., Kent, 1990) states that for any spherical, stationary, non-rotating system with constant mass-to-light ratio $\mathcal{L} = M/L$,

$$2\pi \int_0^\infty I(R)\sigma_{los}^2(R) R\, dR = \frac{GM^2}{3\mathcal{L} r_G} = \frac{M\sigma_V^2}{3\mathcal{L}}, \qquad (1)$$

where $R$ is the projected radius. In (1) the potential energy $W \equiv -GM^2/r_G$ is expressed using the gravitational radius of the system, $r_G$, and the last identity is just the second order Scalar Virial Theorem, where the kinetic energy is written using the (three-dimensional) virial velocity dispersion, $\sigma_V$. In words, *the integral of the l.o.s. velocity dispersion, weighted on the surface brightness profile of a spherical non-rotating galaxy with constant mass-to-light ratio, is equal to 1/3 of its (three-dimensional) virial velocity dispersion.* The proof of (1) can be obtained using for $I\sigma_{los}^2$ the integral expression given in equation (4.61) of Binney and Tremaine (1987) (hereafter BT87, p.208), changing the order of integration, and integrating by parts using the Jeans' equation in spherical coordinates, as shown for example by Tonry (1983). The many steps of the proof can obscure the simplicity of the result, whose importance is intuitively clear: the potential energy is related to the total mass of the galaxy via the knowledge of $r_G$, which in principle, for a one-component galaxy (i.e. with no dark matter), can be derived by deprojecting the surface brightness profile. Because of its importance, it is interesting to extend this result to systems of general shape, in order to obtain a major insight in the subject and to generalize (1). For instance Kent (1990) and Merrifield and Kent (1990) derived higher order PVT for spherical systems, but the calculations are increasingly complicate.

In this paper a general approach to this problem is presented introducing the concept of tomography of a collisionless stellar systems of unspecified shape. In Section 2 the main properties of the projection operator are elucidated. In Section 3 the $N$-th order PVT for systems of any shape is obtained, and in Section 4 the concept of tomography is presented, together with a theorem on the angular mean of the $N$-th power of the velocity component along the line-of-sight direction. Finally in Section 5 the tomography is applied to the generalized PVT's; in the particular case of spherical symmetry the family of any order PVT's [including (1)] is recovered, without using explicit expressions for the l.o.s. velocity dispersion, or spherical coordinates.

## 2　　The projection operator

Let us consider a one-component collisionless stellar system, described in an orthogonal reference system $S = (O; \mathbf{e}_1, \mathbf{e}_2, \mathbf{e}_3)$ by its density $\rho(\mathbf{x}; t) = \mathcal{L}\nu(\mathbf{x}; t)$, where $\nu$ is the density of light. In $S$ we introduce the orthogonal observer's system $S' = (O'; \mathbf{f}_1, \mathbf{f}_2, \mathbf{f}_3)$, and we assume that $O \equiv O'$ at any time, and that $S$ and $S'$ are inertial. Throughout the paper the repeated indices sum convention is used, so that $\mathbf{x} = x_i \mathbf{e}_i$ is a vector in $S$ and $\boldsymbol{\xi} = \xi_j \mathbf{f}_j$ is the same vector in $S'$. Following Binney (1985) we do not use the usual Euler angles to specify the orientation of $S'$ with respect to $S$ but instead we apply a 3-2-3 rotation[1]. In this way $\vartheta$ and $\varphi$ coincide with the polar coordinates of $\mathbf{f}_3$ in $S$, considering $\mathbf{e}_3$ to be the polar axis and $\mathbf{e}_1$ the azimuthal one:

$$\boldsymbol{\xi} = \mathbf{R}\mathbf{x}, \tag{2}$$

with $\mathbf{R} = \mathbf{R}_3(\psi)\mathbf{R}_2(\vartheta)\mathbf{R}_3(\varphi)$. Then the unit vector $\mathbf{n} = \mathbf{f}_3$ from $O$ to the observer is given by

$$\mathbf{n}^T = (\cos\varphi\sin\vartheta,\ \sin\varphi\sin\vartheta,\ \cos\vartheta). \tag{3}$$

Every observed property $\overline{F}_{los}$ associated with the light distribution of a stellar system is the projection onto the plane $\pi' = (\xi_1, \xi_2) \perp \mathbf{n}$ of $\overline{F}(\mathbf{x}; t)$ (see Sect. 3), weighted with the system light density:

$$I(\xi_1, \xi_2; t)\overline{F}_{los}(\xi_1, \xi_2; t) = \int_{\Re} \nu(\mathbf{R}^T\boldsymbol{\xi}; t)\overline{F}(\mathbf{R}^T\boldsymbol{\xi}; t)\, d\xi_3. \tag{4}$$

$\mathbf{R}$ being orthogonal, the projection operator satisfies the following relation:

$$\int_{\Re^2} I(\xi_1, \xi_2; t)\overline{F}_{los}(\xi_1, \xi_2; t)\, d\xi_1\, d\xi_2 = \int_{\Re^3} \nu(\mathbf{x}; t)\overline{F}(\mathbf{x}; t)\, d^3\mathbf{x}. \tag{5}$$

---

[1] Starting from two coincident systems, the Euler angles are defined by a rotation around $\mathbf{e}_3$ ($\varphi$), then around $\mathbf{e}_1'$ ($\vartheta$) and finally around $\mathbf{e}_3''$ ($\psi$). In this 3-1-3 rotation the angles $\varphi$ and $\vartheta$ are not the usual spherical angular coordinates of $\mathbf{e}_3''$ in $S$.

For example, the surface integral of $I$ extended over $\pi'$ equals $L$.

## 3 The Projected Virial Theorems

The definitions introduced up to now are all *geometrical*, i.e. they do not involve the dynamics of the system. All the *dynamical* properties of a collisionless stellar system are known if its distribution function $f = f(\mathbf{x}, \mathbf{v}; t)$ is known. This function obeys the Collisionless Boltzmann Equation (see, e.g., BT87, p.192):

$$\frac{\partial f}{\partial t} + v_i \frac{\partial f}{\partial x_i} - \frac{\partial \phi}{\partial x_i} \frac{\partial f}{\partial v_i} = 0. \tag{6}$$

In fact every *macroscopic* function $\overline{F}(\mathbf{x}; t)$ (i.e. describing some property of a small element of the system but containing a large number of particles) is defined by integrating on the velocity space its *microscopic* counterpart $F(\mathbf{v})$, using $f$ as weight function:

$$\rho(\mathbf{x}; t)\overline{F}(\mathbf{x}; t) \equiv \int_{\Re^3} f(\mathbf{x}, \mathbf{v}; t) F(\mathbf{v}) \, d^3\mathbf{v}. \tag{7}$$

Integrating (6) over $\Re^3$ after multiplication by $F$, and considering (7) one obtains:

$$\frac{\partial \rho \overline{F}}{\partial t} + \frac{\partial \rho \overline{v_i F}}{\partial x_i} + \rho \overline{\left(\frac{\partial F}{\partial v_i}\right)} \frac{\partial \phi}{\partial x_i} = 0. \tag{8}$$

Since we are looking for a generalization of the classical PVT, where the second order monomials made with the velocity components are used, we assume for $F$ the $N$-th power of the $\mathbf{v}$ component along $\mathbf{n}$:

$$F(\mathbf{v}) = <\mathbf{n}, \mathbf{v}>^N \equiv v^N, \tag{9}$$

where $<\,,\,>$ is the inner product in the euclidean space $\Re^3$, and $N \geq 0$ is an integer. For example, for $N = 1$

$$\overline{v}(\mathbf{n}, \mathbf{x}; t) = \overline{<\mathbf{n}, \mathbf{v}>} = <\mathbf{n}, \overline{\mathbf{v}}(\mathbf{x}; t)>, \tag{10}$$

and for $N = 2$, defining the second order velocity dispersion tensor to be $\sigma_{ij}^2 = \overline{(v_i - \overline{v}_i)(v_j - \overline{v}_j)}$:

$$\overline{v^2}(\mathbf{n}, \mathbf{x}; t) = \overline{v}^2(\mathbf{n}, \mathbf{x}; t) + \sigma_p^2(\mathbf{n}, \mathbf{x}; t), \tag{11}$$

where $\sigma_p^2 = n_i n_j \sigma_{ij}^2$. The expansion of (9) for $N \geq 0$ is:

$$v^N = \sum_{K=0}^{N} \sum_{L=0}^{K} \binom{N}{K}\binom{K}{L} (n_3^{N-K} n_1^{K-L} n_2^L)(v_3^{N-K} v_1^{K-L} v_2^L). \tag{12}$$

It is evident that the addenda in $v^N$ (and so in $\overline{v^N}$) are the products between the l.o.s. independent velocity monomials and the angular coefficients that on the contrary depend only on **n**. Projecting $\overline{v^N}$ one obtains

$$I(\xi_1,\xi_2;t)\overline{v^N}_{los}(\xi_1,\xi_2;t) = \int_{\Re} \nu\overline{v^N}\, d\xi_3, \tag{13}$$

and integrating the l.h.s. of (13) over $\pi'$, one finds from (5)

$$\mathcal{L}\int_{\Re^2} I(\xi_1,\xi_2;t)\overline{v^N}_{los}(\xi_1,\xi_2;t)\, d\xi_1\, d\xi_2 = \int_{\Re^3} \rho\overline{v^N} d^3\mathbf{x} \equiv 2K_{N,los} \tag{14}$$

where $K_{N,los}$ is the projected *kinetic energy* of order $N$, which depends on the direction of the l.o.s. through the **n** angular coordinates. For example, for $N=1$ this quantity is the projection of the total momentum (and is 0 according to the total momentum conservation), and for $N=2$ it is the usual projected kinetic energy. From (8), the equation for $\overline{v^N}$ which is necessary to derive the generalized PVT is:

$$\frac{\partial \rho\overline{v^N}}{\partial t} + \frac{\partial \rho\overline{v_i v^N}}{\partial x_i} + Nn_i\rho\overline{v^{N-1}}\frac{\partial \phi}{\partial x_i} = 0. \tag{15}$$

In the specific case of a stationary system, after multiplication of (15) by $x_k$ and integration on $\Re^3$ one obtains three equations. Multiplying each of these equations for $n_k$, and summing, we finally obtain the $N$-th order PVT:

**Theorem** *For $N \geq 0$,*

$$2K_{N+1,los} = -Nn_in_kW_{N-1,ik} = -NW_{N-1,los} \tag{16}$$

*where*

$$W_{N-1,ik} \equiv -\int_{\Re^3} \rho\overline{v^{N-1}}x_k\frac{\partial \phi}{\partial x_i}\, d^3\mathbf{x} = W_{N-1,ki}. \tag{17}$$

For $N=1$, the r.h.s of (16) do not depends on velocity, $W_{0,ik}$ being the usual potential energy tensor (see, e.g., BT87, p. 67), and one obtains the classical second order PVT.

## 4  The angular mean operator

Let us suppose that the observer can move all around the system, observing some projected property $\overline{F}_{los}$ that generally will depend on the l.o.s. direction. This dependence can be eliminated calculating the angular mean of $\overline{F}_{los}$ over the solid angle:

$$[\overline{F}_{los}]_\Omega \equiv \frac{1}{4\pi}\int_{4\pi} \overline{F}_{los}\, d^2\mathbf{\Omega}, \tag{18}$$

where now $[\overline{F}_{los}]_\Omega$ only depends on the intrinsic (3-dimensional) properties of the system. We will call this procedure *tomography*. For spherically symmetric system the result clearly coincides with the projected property along any l.o.s. direction. Then if we succeeded in the explicit calculation of (18), we have at the same time (a) a generalization for systems of unspecified shape of the known results for spherically symmetric systems and (b) a proof of the relations holding for spherically symmetric systems avoiding use of particular algebraic expressions between the quantities to be projected (see, e.g. BT87, p. 205; Tonry, 1983; Merrifield and Kent, 1990; Kent, 1990). Indeed using the *tomography* the calculations are done using only cartesian coordinates. In our particular case, due to the linearity of the $v^N$ with respect to the angular coefficients and velocity monomials [see (12)], it is evident that:

$$[\overline{v^N}]_\Omega = \overline{[v^N]_\Omega}. \tag{19}$$

So we can equally use $v^N$ or $\overline{v^N}$ in any discussion involving the angular mean operator. In order to evaluate (19) for $N \geq 0$, we need the following

**Lemma** *For any integers $N \geq K \geq L \geq 0$ then:*

$$\binom{N}{K}\binom{K}{L} [n_3^{N-K} n_1^{K-L} n_2^L]_\Omega = \frac{\Gamma[(N+2)/2][1+(-1)^{N-K}][1+(-1)^{K-L}][1+(-1)^L]}{8(N+1)\Gamma[(N-K+2)/2]\,\Gamma[(K-L+2)/2]\,\Gamma[(L+2)/2]}. \tag{20}$$

**Proof.** The angular mean is calculated using formula 8.380.2 of Gradstheyn and Ryzhik (1965). Then, using $\binom{a}{b} = \Gamma(a+1)/[\Gamma(b+1)\Gamma(a+b-1)]$ and the duplication formula for $\Gamma$, after some algebra the result is proved.

This gives the following:

**Theorem** *For any integer $N = 2n+1 > 0$,*

$$[v^N]_\Omega = 0; \tag{21}$$

*for any integer $N = 2n \geq 0$,*

$$[v^N]_\Omega = \frac{1}{1+2n} \sum_{k=0}^{n} \sum_{l=0}^{k} \binom{n}{k}\binom{k}{l} (v_3^{n-k} v_1^{k-l} v_2^l)^2. \tag{22}$$

**Proof.** From (12) it follows that in each addendum of $v^N$ the sum of the exponents is equal to $N$, and from (20) only the terms with three even exponents have angular mean different from zero. For $N = 2n+1$ at least one among $N-K$, $K-L$ and $L$ is odd. For $N = 2n$ the numbers $N-K$, $K-L$, and $L$, are necessarily either all even, or one even and two odd: setting $N = 2n$, $K = 2k$, $L = 2l$, also (22) is proved.

For example, for $N = 2$, $[v^2]_\Omega = (v_1^2 + v_2^2 + v_3^2)/3$, and from (11)

$$\overline{[v^2]_\Omega} = [\overline{v^2}]_\Omega = \frac{\overline{v_1}^2 + \overline{v_2}^2 + \overline{v_3}^2}{3} + \frac{\sigma_{11}^2 + \sigma_{22}^2 + \sigma_{33}^2}{3}. \tag{23}$$

In particular the identity $[\sigma_p^2]_\Omega = (\sigma_{11}^2 + \sigma_{22}^2 + \sigma_{33}^2)/3$ is proved. The analogue of (15), i.e. the equation for the angular mean value $[\overline{v^N}]_\Omega$ is immediately obtained:

$$\frac{\partial \rho \overline{[v^N]_\Omega}}{\partial t} + \frac{\partial \rho \overline{v_i [v^N]_\Omega}}{\partial x_i} + \rho \overline{\left(\frac{\partial [v^N]_\Omega}{\partial v_i}\right)} \frac{\partial \phi}{\partial x_i} = 0. \qquad (24)$$

## 5   Tomography of the general PVT

We can now obtain the general expression of the angular mean value of the PVT of any order for stellar systems of unspecified shape from (16). The angular mean value of the projected kinetic energy is obtained from (14) using (21) and (22). The mean value of the generalized gravitational energy in (25) is apparently more complicate. On the contrary it is immediately obtained from (17) observing that

$$N [n_i n_k \overline{v^{N-1}}]_\Omega = \frac{1}{N+1} \overline{\frac{\partial^2 [v^{N+1}]_\Omega}{\partial v_i \partial v_k}}, \qquad (25)$$

and using (21) and (22). For example, the tomography of the classical PVT is given setting $N = 1$ in (16):

$$2[K_{2,los}]_\Omega = -[W_{0,los}]_\Omega = -\frac{W}{3}. \qquad (26)$$

Considering now a spherically symmetric system *the integral of the l.o.s. square velocity weighted on I divided by $\mathcal{L}$ is equal to 2/3 of the total kinetic energy of the system*; moreover, if the system is characterized by pure velocity dispersion, (1) is naturally recovered.

## 6   Acknowledgements

I would like to thank the referee for comments that greatly improved the manuscript, and Prof. G. Contopoulos and Prof. J. Henrard for their encouragements. Sixty per cent of this work was supported by Italian Ministry of Research (MURST).

## 7   References

Binney, J.:1985 'Testing for triaxiality with kinematic data', *Mon. Not. R. Ast. Soc.* **212**, 767-781

Binney, J. and Tremaine, S.:1987 *Galactic Dynamics*, Princeton University Press, Princeton NJ (BT87).

Ciotti, L. and Pellegrini, S.:1992 'Self–Consistent Two–Component Models of Elliptical Galaxies', *Mon. Not. R. Ast. Soc.* **255**, 561-571

Gradshteyn, I.S. and Ryzhik, I.M.: 1965 *Tables of Integrals Series and Products*, (London:Academic Press).

Kent, S.M.:1990 'New Virial Theorems for Spherically Symmetric Systems', *Mon. Not. R. Ast. Soc.* **247**, 702-704.

Merrifield, R.M. and Kent, S.M.:1990 'Fourth Moments and the Dynamics of Spherical Systems', *Astron. J.* **99**, 1548-1557.

Tonry, J.L.:1983 'Anisotropic Velocity Dispersions in Spherical Galaxies', *Astrophys. J.* **266**, 58-68.